\documentclass[aps,prl,twocolumn,superscriptaddress,showpacs]{revtex4}

\usepackage{graphicx}% Include figure files
\usepackage{dcolumn}% Align table columns on decimal point
\usepackage{bm}% bold math
\begin{document}
\def \FeTeSe{Fe$_{1+\delta}$Te$_{1-x}$Se$_x$}
\def \FeTe{Fe$_{1+\delta}$Te}
\def \Ba122{BaFe$_2$As$_2$}
\def \Ca122{CaFe$_2$As$_2$}
\def \CoBa122{Ba(Fe$_{1-x}$Co$_x$)$_2$As$_2$}
\def \KBa122{Ba$_{1-x}$K$_x$Fe$_2$As$_2$}
\def \LSCO{La$_{2-x}$Sr$_x$CuO$_4$}
\def \Sr327{Sr$_3$Ru$_2$O$_7$}
\def \Tc{$T_c$}
\def \Ts{$T_{s}$}
\def \TN{$T_N$}
\def \AF{antiferromagnetic}

\title{Divergent nematic susceptibility in an iron arsenide superconductor}

% repeat the \author .. \affiliation  etc. as needed
% \email, \thanks, \homepage, \altaffiliation all apply to the current
% author. Explanatory text should go in the []'s, actual e-mail
% address or url should go in the {}'s for \email and \homepage.
% Please use the appropriate macro foreach each type of information

% \affiliation command applies to all authors since the last
% \affiliation command. The \affiliation command should follow the
% other information
% \affiliation can be followed by \email, \homepage, \thanks as well.

%\email[]{Your e-mail address}
%\homepage[]{Your web page}
%\thanks{}
%\altaffiliation{}
\author{Jiun-Haw Chu}
\affiliation{Department of Applied Physics and Geballe Laboratory for Advanced Materials, Stanford University, Stanford, California 94305, USA}
\affiliation{Stanford Institute of Energy and Materials Science, SLAC National Accelerator Laboratory, 2575 Sand Hill Road, Menlo Park 94025,California 94305, USA}
\author{Hsueh-Hui Kuo}
\affiliation{Department of Materials Science and Engineering and Geballe Laboratory for Advanced Materials, Stanford University, Stanford, California 94305, USA}
\affiliation{Stanford Institute of Energy and Materials Science, SLAC National Accelerator Laboratory, 2575 Sand Hill Road, Menlo Park 94025,California 94305, USA}
\author{James G. Analytis}
\affiliation{Department of Applied Physics and Geballe Laboratory for Advanced Materials, Stanford University, Stanford, California 94305, USA}
\affiliation{Stanford Institute of Energy and Materials Science, SLAC National Accelerator Laboratory, 2575 Sand Hill Road, Menlo Park 94025,California 94305, USA}
\author{Ian R. Fisher}
\affiliation{Department of Applied Physics and Geballe Laboratory for Advanced Materials, Stanford University, Stanford, California 94305, USA}
\affiliation{Stanford Institute of Energy and Materials Science, SLAC National Accelerator Laboratory, 2575 Sand Hill Road, Menlo Park 94025,California 94305, USA}

%Collaboration name if desired (requires use of superscriptaddress
%option in \documentclass). \noaffiliation is required (may also be
%used with the \author command).
%\collaboration can be followed by \email, \homepage, \thanks as well.
%\collaboration{}
%\noaffiliation

\date{\today}

\begin{abstract}
Within the Landau paradigm of continuous phase transitions, ordered states of matter are characterized by a broken symmetry. Although the broken symmetry is usually evident, determining the driving force behind the phase transition is often a more subtle matter due to coupling between otherwise distinct order parameters. In this paper we show how measurement of the divergent nematic susceptibility of an iron pnictide superconductor unambiguously distinguishes an electronic nematic phase transition from a simple ferroelastic distortion. These measurements also reveal an electronic nematic quantum phase transition at the composition with optimal superconducting transition temperature.
\end{abstract}

% insert suggested PACS numbers in braces on next line
% insert suggested keywords - APS authors don't need to do this
%\keywords{}
%\maketitle must follow title, authors, abstract, \pacs, and \keywords
\maketitle

% body of paper here - Use proper section commands
% References should be done using the \cite, \ref, and \label commands
An electronic nematic phase transition refers to a phase transition in which the electronic system self-organizes with orientational order without forming spatial periodic order\cite{Fradkin_2010}. For crystalline systems, such a nematic transition breaks a discrete rotational symmetry of the crystal lattice without altering the existing translational symmetry. Canonical examples include half filling Quantum Hall states\cite{Lilly_1999}, the field induced metamagnetic state in \Sr327\cite{Borzi_2007} and most recently the hidden order state in URu$_2$Si$_2$\cite{Okazaki_2011}, in which cases the crystal lattice's fourfold symmetry remains almost unperturbed yet the electronic ground states exhibit a strong two-fold anisotropy.\cite{Stingl_2011}. Recently both high \Tc\ cuprates\cite{Ando_2001,Hinkov_2008,Daou_2010,Lawler_2010} and iron pnictides\cite{Chuang_2010,Chu_2010,Fernandes_2010} have been proposed as candidate platforms that might harbour an electronic nematic phase, which opens up exciting new possibilities related to the interplay of nematic order with high temperature superconductivity. However, one of the key doubts accompanied by the experimental discoveries is that the crystal lattice of these two systems does not retain a fourfold symmetry. In particular, in iron pnictides there is an orthorhombic structural distortion accompanying the rapid increase of resistivity anisotropy, which puts the legitimacy of the term "electronic nematic" into question. Here we report measurements of the resistivity anisotropy of \CoBa122\ induced by a tunable uni-axial strain, which exhibits a divergent behaviour as the system approaches the phase transition from the high temperature side. Our result explicitly shows that the structural phase transition in \CoBa122\ is purely driven by the instability in the electronic part of the free energy, and furthermore reveals an electronic nematic quantum phase transition at the composition with optimal superconducting transition temperature.

\begin{figure}
\includegraphics[width=8.5cm]{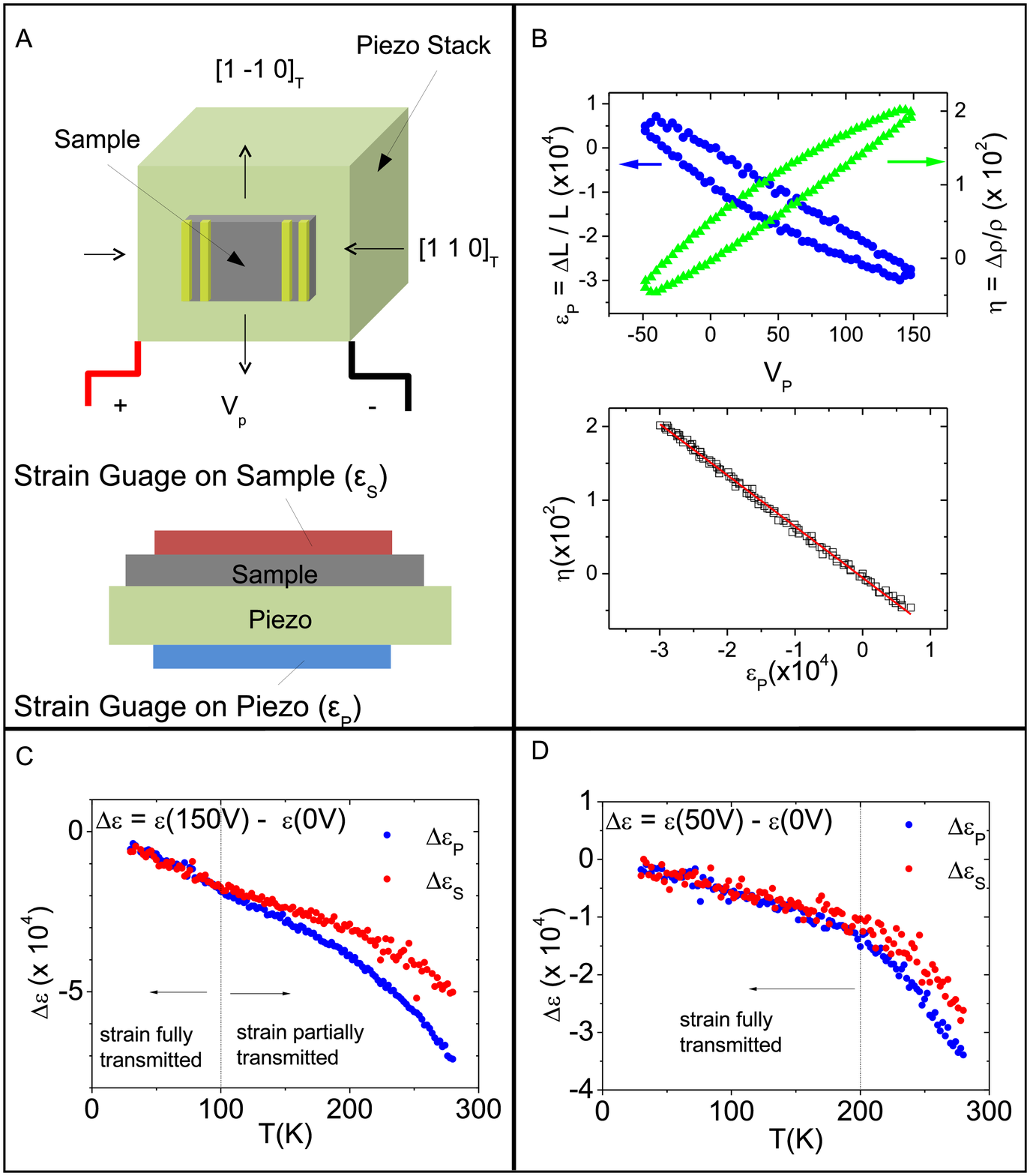}
\caption{\label{Fig:Fig1} (Color online)\textbf{(A)} Schematic diagram of a piezoresistance measurement (upper panel) and of a strain gauge measurement(lower panel). Details about the configuration are described in the supporting online materials. \textbf{(B)}(Upper) The relative change of resistivity ($\eta = \Delta\rho/\rho_0$) of a \Ba122 sample and the strain measured by strain gauge on piezo ($\epsilon_P = \Delta L/L$) as a function of voltage at $T = 140K$. The strain and resistance were measured along the [1 1 0]$_T$ direction of the crystal. (Lower) Same set of data but $\eta$ is plotted against $\epsilon_P$ . The red line is a linear fit to the data. \textbf{(C)(D)} The difference of the strain between zero applied voltage and (C) $V_p  = 150V$ and (D) $V_p  = 50V$. Dash lines indicate the temperatures below which the strain is fully transmitted to the sample. For low voltage this temperature window spans well above \Ts\ for all compositions studied.}
\end{figure}

We apply a tuneable in-plane uniaxial strain to single crystal samples of \CoBa122\ to probe the nematic response. As shown in Fig. \ref{Fig:Fig1}(A), by gluing the sample on the side wall of a piezostack, strains can be applied by the deformation of the piezo, which is controlled by an applied voltage(V$_P$)\cite{Shayegan_2003}. The strain (i.e. the fractional change of length along the current direction, $\epsilon_P = \Delta L/L$) was monitored via a strain gauge glued on the back side of the piezo stack. Both $\epsilon_P$ and the fractional change of resistivity ($\eta = \Delta\rho/\rho_0$, where $\rho_0$ is the resistivity of the free standing sample before gluing on the piezo stack) were measured at constant temperature while the applied voltage was swept, as shown in Fig. \ref{Fig:Fig1}(B). The voltage dependence of $\eta$ and $\epsilon_P$ shows hysteretic behaviour due to the ferroelectric nature of the piezo materials, yet the two quantities exhibit a linear relationship without any hysteresis(Fig. \ref{Fig:Fig1}(B)). The negative slope of $\eta(\epsilon_P)$ indicates that the resistivity is higher along the shorter bonding direction, consistent with previous results\cite{Chu_2010,Tanatar_2010,Fisher_2011}.

The amount of strain transmitted to the sample ($\epsilon_S$) can be assessed by gluing another strain gauge on the top surface of the crystal, shown schematically in the lower panel of Fig. \ref{Fig:Fig1}(A). The comparison of $\epsilon_S$ and $\epsilon_P$ for a Ba(Fe$_{0.955}$Co$_{0.045}$)$_2$As$_2$ sample is summarized in Fig \ref{Fig:Fig1} (C and D).  For applied voltages $|V_p| <150V$, the strain is fully transmitted to the sample for temperatures below approximately 100 K for typical thickness crystals (less than 100 $\mu$m). For lower voltages, $|V_p| <50V$, the strain is fully transmitted to even higher temperatures (Figure 1(D)). The maximum strain that can be applied ($|\epsilon|<5\times 10^{-4}$) is substantially less than the lattice distortion developed below the phase transition($10^{-2}\sim 10^{-3}$), and, as we show below, the system is always in the regime of linear response.

The induced fractional change of the resistivity $\eta$ provides a direct measure of the electronic nematic order parameter. Specifically, in the ordered state the resistivity anisotropy $\psi = (\rho_b-\rho_a)/(\rho_b+\rho_a)$ can be used to define a nematic order parameter. For a strained crystal in the tetragonal state $\rho_b$ and $\rho_a$ refer to the resistivity in directions parallel and perpendicular to the applied compressive stress. It can be easily shown that $\eta = \psi$ if the increase in $\rho_b$ equals the decrease in $\rho_a$, and that the two quantities are directly proportional even if this is not the case. The same is also true for the derivatives of these quantities such that $d\eta/d\epsilon\propto d\psi/d\epsilon$ (supplementary online text).

\begin{figure}
\includegraphics[width=8.5cm]{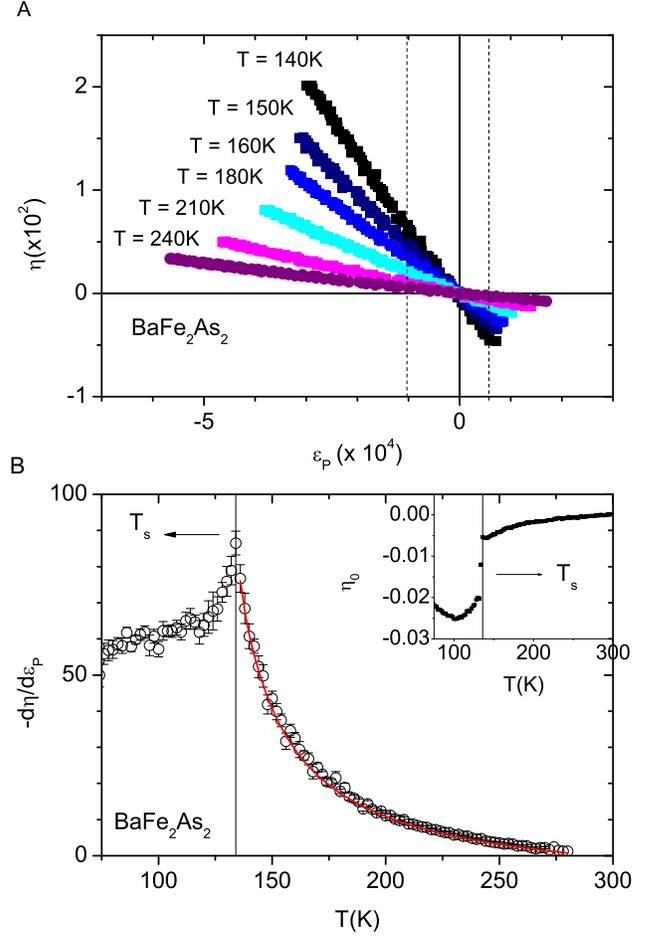}
\caption{\label{Fig:Fig2} \textbf{(A)} Representative data for \Ba122 showing the relative change of resistivity ($\eta = \Delta\rho/\rho_0$) as a function of strain ($\epsilon_P = \Delta L/L$) at several temperatures above \Ts . The nematic response was obtained by a linear fit of the data near zero applied voltage ( $-5\times 10^{-5} < \epsilon_p(V) - \epsilon_p(0) < 1\times 10^{-4}$, indicated by the vertical dashed lines). \textbf{(B)} Temperature dependence of the nematic response $d\eta/d\epsilon_P$. Vertical line indicates the structural transition temperature \Ts = 138$K$. Red line shows fit to mean field model, described in the main text.}
\end{figure}

Representative data showing the electronic nematicity ($\eta$) as a function of strain ($\epsilon_P$) for a \Ba122\ sample are shown in Fig. \ref{Fig:Fig2}(A) at various temperatures above the structural transition temperature \Ts\ . Data were fit by a straight line in a small range of strain near zero applied voltage. As shown in Fig. \ref{Fig:Fig2}(D), the quantity $d\eta/d\epsilon$, which essentially measures the nematic response induced by a constant strain, diverges upon approaching \Ts\ from above. This divergent behavior is reminiscent of the resistivity anisotropy observed above \Ts\ for samples held in a mechanical clamp\cite{Chu_2010}. However, as we explain below, there is an important distinction between measurements made under condition of constant stress(mechanical clamp), and constant strain(measurement of $d\eta/d\epsilon$ in the current set up.)

From the thermodynamic point of view the stress and strain are conjugate variables, and the stress (here denoted as $h$) is the externally controllable force, whereas strain is the response of a mechanical system. Intuitively it might be more reasonable to regard stress as a symmetry breaking field. However from the electron nematic stand point, stress only couples indirectly to the nematic order parameter through strain. This relationship can be best understood by the following Ginzburg-Landau free energy:
\begin{eqnarray}\label{Eq:GLfree}
F = \frac{a}{2}\psi^2 + \frac{b}{4}\psi^4 + \frac{c}{2}\epsilon^2 + \frac{d}{4}\epsilon^4 - \lambda\psi\epsilon -h\epsilon
\end{eqnarray}
Here $\psi$ represents the electronic nematic order parameter, measured by the resistivity as discussed above, $\epsilon$ is the elastic strain, and $h$ is its conjugate stress. $a$, $b$, $c$ and $d$ are the coefficients of the two order parameters in the usual power series expansion, and $\lambda$ is the coupling constant. If there is a phase transition driven by the electronic degree of freedom, then the coefficient $a$ becomes zero at some temperature, i.e. $a = a_0(T- T^*)$, whereas the other coefficients are temperature independent. The bilinear coupling term $\lambda\epsilon\psi$ renormalizes the coefficient of $\epsilon^2$ ($c$ becomes $c - \lambda^2/(a_0(T- T^*))$) such that the crystal lattice distorts simultaneously with the onset of nematic order.  On the other hand if the phase transition is due to a structural instability, then it is the coefficient $c$ that becomes zero ($c = c_0(T - T^*)$) \cite{Cano}. Therefore the driving force can be distinguished by determining the temperature dependence of the bare $a$ and $c$ coefficients.

With this in mind we can now ask what is the difference between measuring the response of electronic nematicity $\psi$ under constant strain $\epsilon$ rather constant stress $h$. This can be answered explicitly by calculating the quantities of $d\psi/dh$ and $d\psi/d\epsilon$ under the constraint of minimizing the free energy(supplementary online text):
\begin{eqnarray}\label{Eq:ChiStress}
\frac{d\psi}{dh} = \frac{\lambda}{ac - \lambda^2}
\end{eqnarray}
\begin{eqnarray}\label{Eq:ChiStrain}
\frac{d\psi}{d\epsilon} = \frac{\lambda}{a}
\end{eqnarray}
From these expressions, it is clear that the nematic response under a constant stress (eq. \ref{Eq:ChiStress}) will show a $1/T$ divergence no matter whether the driving force is a structural or electronic phase transition. However the nematic response under a constant strain will only diverge when it is a true electronic nematic phase transition(eq. \ref{Eq:ChiStrain}). In this sense, the divergence in $d\eta/d\epsilon$ shown in Fig. \ref{Fig:Fig2} is direct evidence that \Ba122\ suffers a true electronic nematic instability, and the structural transition merely passively follows the nematic order. Since strain is a field to the nematic order parameter, we refer to the quantity $d\psi/d\epsilon$ as the nematic susceptibility.

From Eq. \ref{Eq:ChiStrain} $d\psi/d\epsilon = \lambda/a = \lambda/(a_0(T - T^*))$, it is natural to fit the data of $d\eta/d\epsilon$ in Fig, \ref{Fig:Fig2}(B) with a Curie-Weiss temperature dependence. However Eq. \ref{Eq:ChiStrain} is only valid in the limit of vanishing strain, at which one can disregard the higher order non-linear terms. In the realistic experiment situation, there is always some intrinsic built in strain even at zero applied voltage due to the different thermal contraction of the sample and the piezo stack. To take into account this built in-strain, we perform a numerical fit based on the following expression:
\begin{eqnarray}\label{Eq:fit}
\frac{d\eta}{d\epsilon} = \frac{\lambda}{a_0(T - T^*) + 3b\eta_0^2} + \chi_0
\end{eqnarray}

The effect of next order non-linear term is included in the $3b\eta_0^2$ in the denominator, where $\eta_0$ is the resistivity anisotropy induced by the built in strain as a function of temperature, measured by the difference of resistivity of a sample before and after gluing on the piezo stack. In addition to $a_0$ and $b$ introduced before, $\chi_0$ is a fitting parameter to model the intrinsic piezoresistivity effect of the materials that is unrelated to the electron nematic phase transition.

The result of this fitting is plotted in Fig.\ref{Fig:Fig2}(B) as a solid red curve, which is in excellent agreement with measured data $d\eta/d\epsilon$. Interestingly, the mean field critical temperature $T^*$ obtained from the fitting is 116$K$, 22$K$ lower than the actual phase transition temperature (\Ts$=138K$ ). This can also be understood from the Ginzburg-Landau free energy in Eq. \ref{Eq:GLfree}. By minimizing the free energy it can be derived that the nonzero nematic and structural order parameters onset simultaneously at a temperature $T_S = T^* + \lambda^2/(a_0c)$, higher than $T^*$. This is due to the bilinear coupling between the electronic nematic system and the crystal lattice, which lifts the critical temperature of the electronic instability to a higher temperature. Physically, the lattice provides a polarizable medium, which enhances the nematic instability.

\begin{figure}
\includegraphics[width=8.5cm]{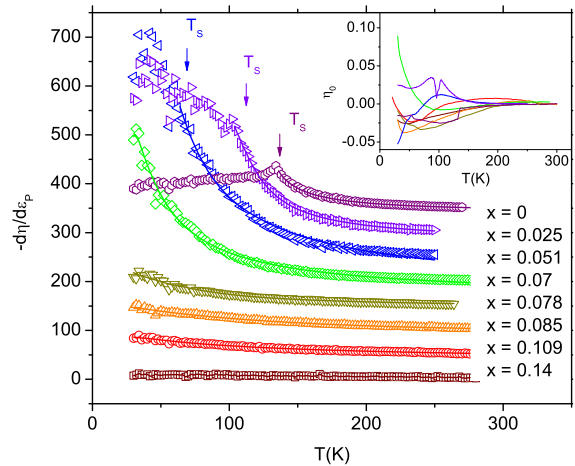}
\caption{\label{Fig:Fig3} Temperature dependence of the nematic susceptibility of \CoBa122\ for various compositions(open symbols). Successive data sets are offset vertically by 50 for clarity. Solid lines are fits based on a phenomenological Ginzburg-Landau theory, taking into account an intrinsic built-in strain(supplementary online text)}
\end{figure}

\begin{figure}
\includegraphics[width=8.5cm]{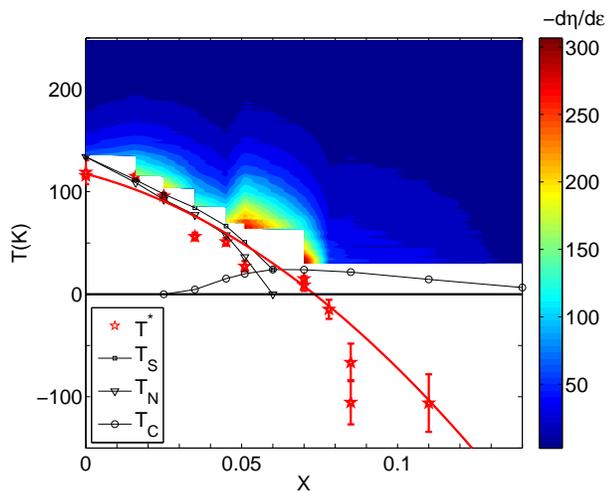}
\caption{\label{Fig:Fig4} Evolution of the nematic susceptibility ($d\eta/d\epsilon$) of \CoBa122\ as a function of temperature and doping. Structural, magnetic and superconducting transition temperatures (\Ts\ ,\TN\ and \Tc) are shown as squares, triangles and circles. The mean field electronic nematic critical temperature($T^*$) obtained from the fit to the data in Fig. \ref{Fig:Fig3} are shown as open red stars. The evolution of nematic susceptibility and nematic critical temperatures clearly indicates an electronic nematic quantum phase transition occurs close to optimal doping.}
\end{figure}

The divergence of $d\eta/d\epsilon$ not only reveals the tendency towards an electronic nematic phase transition, but also measures the strength of nematic fluctuations, according to the fluctuation-dissipation theorem. We have measured $d\eta/d\epsilon$ of \CoBa122\ samples for doping concentration ranging from the undoped parent compound to overdoped compositions, as shown in Fig.\ref{Fig:Fig3}. The magnitude of $d\eta/d\epsilon$ is plotted as a color-map in the composition verses temperature phase diagram in Fig.\ref{Fig:Fig4}. For the underdoped part of the phase diagram, $d\eta/d\epsilon$ increases rapidly near the structural phase transition boundary. As the doping concentration increases, the intensity of fluctuations increases, and reaches a maximum near optimal doping concentration, where structural and magnetic transitions are fully suppressed. The nematic fluctuations persist to the overdoped regime, eventually decreasing as the superconducting \Tc\ decreases. The associated softening of the sheer modulus has been extensively studied by resonant ultrasound measurements\cite{Fernandes_2010,Yoshizawa_2012}.

To quantitatively track the evolution of nematic fluctuations across the phase diagram, numerical fits to the data were performed for each composition based on Eq.\ref{Eq:fit}. The obtained $T^*$ is also plotted as a function of composition in Fig.\ref{Fig:Fig4}. It can be clearly seen that the mean field nematic critical temperature $T^*$ closely tracks the actual structural transition temperature \Ts\ in the underdoped regime, and is suppressed to zero at the optimal doping. $T^*$ becomes negative as the doping further increases beyond optimal doping, indicating a ``paranematic" state. Significantly, our experimental data and analysis reveal an electronic nematic quantum phase transition for a composition close to optimal doping. It remains to be seen whether fluctuations associated with this quantum phase transition play an important role in enhancing $T_c$ in the superconducting phase. Nevertheless the existence of nematic fluctuations across such a wide temperature and doping range suggests that they are a fundamental ingredient to describe the normal state of the system\cite{Fernandes_2010}.

The experiment we introduced here is a methodology to detect the electronic tendency towards rotational symmetry breaking as if there is no coupling to the lattice, and is by no means restricted to the iron pnictides. One can also incorporate this methodology with other experimental probes, which allows us to more generally disentangle the cause and effect in systems for which different degrees of freedom strongly couple. For example, it is still an ongoing debate about the microscopic mechanism of the electron nematic phase transition in pnicitides, which could potentially arise from the spin\cite{Fang_2008, Xu_2008, Yildirim_2009, Hu_2011, Fernandes_2012} or orbital\cite{Kruger_2009, Chen_2009, Lv_2009} degree of freedom. By applying a constant strain and measuring temperature dependence of the orbital response by ARPES\cite{Yi_2011} or optical conductivity\cite{Dusza_2011,Nakajima_2011} and measuring the spin response by neutron scattering , the debate might be resolved.
The authors thank C-.C. Chen, P. Coleman, R. M. Fernandes, S. A. Kivelson, A. Mackenzie and Q. Si for helpful discussions. This work is supported by the DOE, Office of Basic Energy Sciences, under contract no. DE-AC02-76SF00515.

\section{Supplementary Online Text}

\subsection{Materials and methods}
\subsubsection{Crystal growth and transport measurements}
Single crystals of \CoBa122\ were grown from a self flux, as described previously(S1), and the cobalt concentration determined by microprobe analysis. The crystals have a plate-like morphology, with the c-axis perpendicular to the plane of the plates. The in-plane orientation was determined by x-ray diffraction. Electrical contacts were made using silver epoxy to sputtered gold pads, with typical contact resistances of 1 $\sim$ 2 $\Omega$. Resistivity measurements were made using a standard 4-point configuration.

\subsubsection{Piezo device}
\begin{figure}
\includegraphics[width=8.5cm]{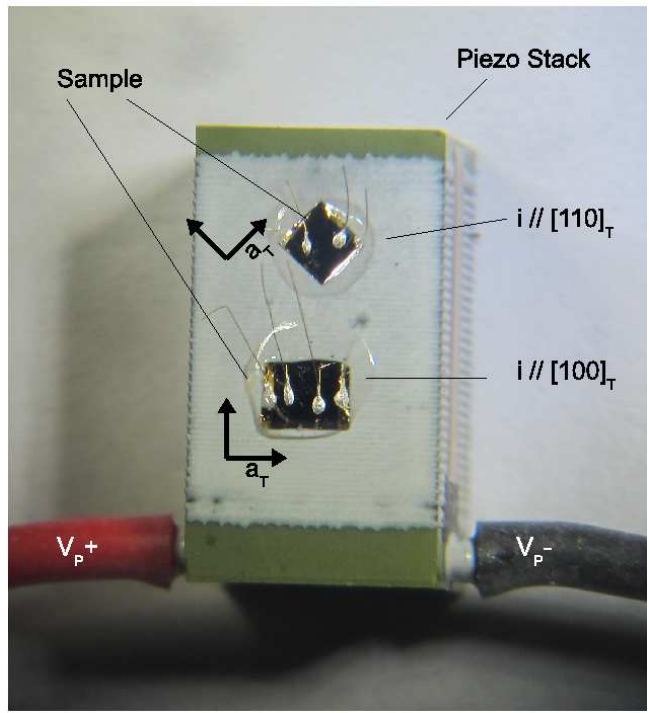}\label{FigS1}
\caption{\label{Fig:S1}Photograph of two representative crystals mounted on a piezo stack used to apply in situ tunable strain. Two samples are mounted, corresponding to $\epsilon//$ [110]$_T$ (upper crystal) and $\epsilon//$ [100]$_T$ (lower crystal). Red/black wires are the positive/negative voltage leads of the piezo. The (0 0 1) surface of the crystal is exposed, enabling transport measurements.}
\end{figure}

\begin{figure}
\includegraphics[width=8.5cm]{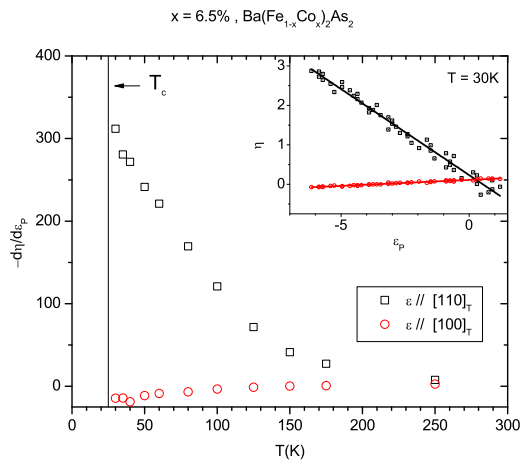}\label{FigS2}
\caption{\label{Fig:S2}Representative data showing the temperature dependence of the nematic response $d\eta/d\epsilon_P$ of Ba(Fe$_{0.935}$Co$_{0.065}$)$_2$As$_2$ samples, for the two configurations shown in fig. S1. Samples were prepared so that the strain and current were aligned along the [1 1 0]$_T$ (Fe-Fe bonding direction) and [1 0 0]$_T$ (Fe-As bonding direction), shown as black and red symbols respectively.  Vertical line indicates the superconducting critical temperature \Tc = 25$K$ of these optimal doped samples. Inset shows the relative change of resistivity ($\eta = \Delta\rho/\rho_0$) as a function of strain ($\epsilon_P = \Delta L/L$) at $T = 30K$. Lines show linear fits.}
\end{figure}

Motivated by previous piezoresistance measurements of quantum Hall systems(S2), we developed a similar scheme to apply in-situ tunable strain for single crystal samples of iron based superconductors. Fig.\ref{Fig:S1} shows a photograph of a representative commercial PZT piezo stack that was used for the experiment(S2). By applying a positive voltage bias through the red and black voltage leads, the piezo stack expands along the long dimension and contracts along the transverse direction. Samples were glued on the side wall using a standard two part epoxy(S3), and a strain gauge was glued on the other side to measure the amount of strain that is applied(S4). The relative orientation of the crystal axes with respect to the piezo stack determines the direction in which the strain is applied. For a control experiment, two samples of \CoBa122\ with $x = 0.065$ (i.e. nearly optimally doped) were measured, one with strain aligned along the tetragonal [110] direction($\epsilon//$[110]$_T$) and one aligned along the tetragonal [100] direction($\epsilon//$[100]$_T$). The resistivity was always measured along the strain direction.  As shown in Fig. \ref{Fig:S2}, for $\epsilon//$[110]$_T$ the temperature dependence of $d\eta/d\epsilon$ exhibits a similar divergent behavior to that of the optimal doped sample shown in the main text Fig.\ref{Fig:Fig4} . However, for $\epsilon//$[100]$_T$ (i.e. at 45 degrees to the orthorhombic direction observed for underdoped compositions) $d\eta/d\epsilon$ has an opposite sign and a much smaller value, and exhibits a much weaker temperature dependence. The significant difference of $d\eta/d\epsilon$ obtained between for the two orientations confirms the following analysis in which the divergent behavior is due to the coupling to nematic fluctuations.

\subsection{Resistivity anisotropy and nematic order parameter}

In general, the electronic nematic order parameter can be expressed as a function of the resistivity anisotropy $\psi = (\rho_b-\rho_a)/(\rho_b+\rho_a)$. Taking a Taylor expansion for small values of $\psi$, the leading term is linear, such that $\psi$ is directly proportional to the electronic nematic order parameter when the order parameter is small.  As discussed in the main text, we measure the relative change of the resistivity $\eta = \Delta\rho/\rho_0$ of a strained samples in the tetragonal state. In the absence of strain $\rho_a = \rho_b = \rho_0$ . When an infinitesimal amount of strain $\delta\epsilon$ is applied, the resultant $\rho_a$ and $\rho_b$ (here $b$ represents the compressive strain direction) can be expressed as a linear function of $\delta\epsilon$.
\begin{eqnarray}
\rho_a = \rho_0(1 - \alpha\delta\epsilon)\\
\rho_b = \rho_0(1 + \alpha'\delta\epsilon)\\
\psi = \frac{\alpha+\alpha'}{2}\delta\epsilon + O(\delta\epsilon^2)
\end{eqnarray}

If $\alpha$ equals $\alpha'$, it's straight forward to show that $\psi = \eta$, otherwise another proportional factor needs to be introduced in the relationship between these two quantities. If only the leading order term in the expansion of $\psi$ in terms of $\delta\epsilon$ is considered, we then get:
\begin{eqnarray}
\delta\epsilon < 0 \Rightarrow \psi = \frac{\alpha' + \alpha}{2\alpha'}\eta\\
\delta\epsilon > 0 \Rightarrow \psi = \frac{\alpha' + \alpha}{2\alpha}\eta
\end{eqnarray}
In principle, the resistivity anisotropy will also be affected by acoustic phonon fluctuations associated with the structural transition. However for a $k=0$ transition the wavelength of such fluctuations is comparable to the sample size, and therefore scattering from such fluctuations makes negligible contribution to the resistivity anisotropy except for extremely close to T$_S$(S5).

\subsection{Formalism of phenomenological Ginzburg-Landau theory}
In the presence of a crystal lattice, an electronic nematic phase transition breaks the crystalline point group rotational symmetry. For example in the case of \CoBa122 ,the C$_4$ symmetry is lowered to C$_2$ symmetry. In this particular case we can write down the system's free energy in terms of an expansion of an Ising order parameter $\psi$;
\begin{eqnarray}
F = \frac{a}{2}\psi^2 + \frac{b}{4}\psi^4
\end{eqnarray}
To minimize the free energy one can take the derivative of $F$ with respect to $\psi$ and set it to zero:
\begin{eqnarray}
\frac{\partial F}{\partial\psi} = a\psi +b\psi^3 = 0\label{eq:bare}
\end{eqnarray}
This equation allows a non-zero solution for $a<0$ and $b>0$. Following the usual procedure, we express $a$ by $a_0(T - T^{*})$ and set $b$ as a positive constant to model the onset of a finite order parameter at temperature $T^{*}$. In the presence of electron-lattice coupling, the non-zero order parameter in the electronic degree of freedom will always induce a finite order parameter in the lattice degree of freedom. In the case of C$_4$ to C$_2$ symmetry breaking the lattice order parameter is measured by a orthorhombic strain $\epsilon = (a-b)/(a+b)$. The effect of electron-lattice coupling can be modelled by the following free energy expression:  
\begin{eqnarray}
F = \frac{a}{2}\psi^2 + \frac{b}{4}\psi^4 + \frac{c}{2}\epsilon^2 + \frac{d}{4}\epsilon^4 - \lambda\psi\epsilon
\end{eqnarray}
The term $\lambda\psi\epsilon$ indicates that electronic order parameter and lattice order parameters are bilinearly coupled, which is the lowest order coupling allowed by symmetry. We take the derivative of free energy with respect to $\psi$ and with respect to $\epsilon$, and set them to be zero.
\begin{eqnarray}\label{eq:cond1}
\frac{\partial F}{\partial\psi} = a\psi +b\psi^3 - \lambda\epsilon = 0
\end{eqnarray}
\begin{eqnarray}\label{eq:cond2}
\frac{\partial F}{\partial\epsilon} = c\epsilon +d\epsilon^3 - \lambda\psi = 0
\end{eqnarray}
These two equations determine the electronic and lattice order parameters as a function of temperature under the constraint of minimizing the free energy. Substituting for $\epsilon$ in equation \ref{eq:cond2} using the expression of $\epsilon = (a\psi +b\psi^3)/\lambda$ obtained from eq. \ref{eq:cond1}, we get:
\begin{eqnarray}\label{eq:rmcond1}
(a - \frac{\lambda^2}{c})\psi +(b + \frac{da^3}{c\lambda^2})\psi^3 = 0
\end{eqnarray}
similarly we get
\begin{eqnarray}\label{eq:rmcond2}
(c - \frac{\lambda^2}{a})\epsilon +(d + \frac{bc^3}{a\lambda^2})\epsilon^3 = 0
\end{eqnarray}
Equation \ref{eq:rmcond1} has the same form as equation \ref{eq:bare}, but the coefficients are renormalized due to coupling to the lattice. In essence, the order parameter becomes non-zero at a temperature $T_S$ higher than $T^{*}$, where $T_S = T^{*} + \frac{\lambda^2}{a_0c}$. We can also look at equation \ref{eq:rmcond2}, which describes the lattice part of the free energy. Due to the coupling to the electronic degree of freedom, the coefficient $c$, which is the "bare" elastic modulus, is also renormalized to a new effective modulus $c_{eff} = c - \lambda^2/a = c - \lambda^2/(a_0(T - T^{*}))$. Even though $c$ might always be positive, meaning that there is no instability in the lattice part of the free energy, the effective modulus still goes to zero at $T_S$, indicating the onset of non-zero spontaneous strain $\epsilon$. Notice that the above derivation is completely reciprocal between the electronic and lattice order parameters -- an instability in the lattice degree of freedom will also lead to non-zero electron nematicity below a structural phase transition. Then the question arises, how do we know that the electronic anisotropy in \CoBa122 is due to a real electronic nematic phase transition, rather than a parasitic effect of a pure elastic phase transition?
A Gedankenexperiment one can do is to put the electrons in an infinitely stiff lattice, i.e. tune the parameter $c$ to infinity. In this scenario if there is really an electronic nematic phase transition, then the lattice will always stay C$_4$ symmetric while the electronic system will spontaneously break C$_4$ symmetry at temperature $T^{*}$, which can be measured by the resistivity anisotropy. %The only problem is that such a lattice does not exist and can possibly only be realized by optical lattice (a convenient excuse for condensed matter physicist).
Another approach is to directly measure the bare coefficients $a$, and determine whether the bare nematic coefficient really becomes zero at some temperature $T^{*}$. This measurement can be done by perturbing the system with a symmetry breaking stress, and measuring the quantity $d\psi/d\epsilon$. To obtain this result, we introduce stress $h$, the conjugate field to strain $\epsilon$, into the free energy:
\begin{eqnarray}
F = \frac{a}{2}\psi^2 + \frac{b}{4}\psi^4 + \frac{c}{2}\epsilon^2 + \frac{d}{4}\epsilon^4 - \lambda\psi\epsilon - h\epsilon
\end{eqnarray}
The stress breaks the dual symmetry between $\psi$ and $\epsilon$ in free energy, because $\psi$ does not couple directly to $h$. Minimizing this free energy with respect to $\psi$ and $\epsilon$ we get a similar set of ``equations of motion" for $\psi$ and $\epsilon$:
\begin{eqnarray}
\frac{\partial F}{\partial\psi} = a\psi +b\psi^3 - \lambda\epsilon = 0\label{eq:cond1stress}\\
\frac{\partial F}{\partial\epsilon} = c\epsilon +b\epsilon^3 - \lambda\psi -h = 0\label{eq:cond2stress}
\end{eqnarray}
To see how system responds to stress, we take the derivative of equation \ref{eq:cond1stress} with respect to $h$:
\begin{eqnarray}
a\frac{d\psi}{dh} +3b\psi^2\frac{d\psi}{dh} - \lambda\frac{d\epsilon}{dh} = 0
\end{eqnarray}
From this we get:
\begin{eqnarray}\label{eq:fullexpress}
\frac{d\psi}{d\epsilon} = \frac{d\psi/dh}{d\epsilon/dh} = \frac{\lambda}{a + 3b\psi^2}
\end{eqnarray}
If we only consider the zero stress limit $h\sim 0$, then $\psi\sim 0$, and the above equation reduces to :
\begin{eqnarray}
\frac{d\psi}{d\epsilon} = \frac{d\psi/dh}{d\epsilon/dh} = \frac{\lambda}{a}
\end{eqnarray}
Therefore if there is truly an electronic nematic phase transition, one should observe a Curie-Weiss diverging behavior of the quantity $d\psi/d\epsilon$ with a critical temperature $T^{*}$. As described in the previous section, our measurement of $d\eta/d\epsilon$ is directly proportional to $d\psi/d\epsilon$. The experiment is equivalent to measuring the temperature dependence of the nematic order parameter at constant strain, as can be appreciated from eq. \ref{eq:cond1stress}. Above $T_S$ we can neglect terms proportional to $\psi^3$, leading to $\psi = \lambda\epsilon/a$ and hence $d\psi/d\epsilon = \lambda/a$.
Although it is highly unphysical, if we also allow a direct coupling between mechanical stress and the nematic order parameter and include a term $gh\psi$ in the free energy, we obtain:
\begin{eqnarray}
\frac{d\psi}{d\epsilon} = \frac{cg+\lambda}{a + \lambda g}
\end{eqnarray}
The separation of $c$ and $a$ into numerator and denominator still allows us to distinguish the electronic instability from structural, since $d\psi/d\epsilon$ diverges only when $a$ goes to zero. Physically, this distinction is possible because $d\psi/d\epsilon$ is related to the ratio of two thermodynamic quantities. We refer $d\psi/d\epsilon$ as a nematic susceptibility because in the limit that $g = 0$, the strain $\epsilon$ is the only field to the nematic order parameter.

\subsection{Numerical fits based on phenomenological Ginzburg-Landau theory}

The measured $d\eta/d\epsilon$ was fitted based on the Ginzburg-Landau theory described in the previous section.  Since the difference of thermal contraction between sample and piezo stack will always induce some finite built in strain even when the bias voltage is zero, the assumption of zero stress limit does not hold in the real experiment. Here, data are fit using the full expression for $d\eta/d\epsilon$ taken at finite $\eta$ obtained from Eq. \ref{eq:fullexpress}:
\begin{figure}
\includegraphics[width=8.5cm]{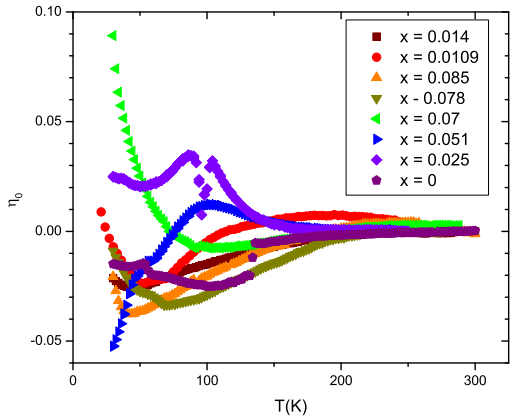}
\caption{ \label{Fig:S3}Temperature dependence of the relative change of resistivity $\eta_0$ induced by the intrinsic built in strain associated with the differential thermal contraction of the piezo stacks and the crystals glued on their surface . Data presented here together with $d\eta/d\epsilon$ presented in the main text Fig. 3 were used for the numerical fit based on Ginzburg-Landau theory.}
\end{figure}

\begin{eqnarray}
\left.\frac{d\eta}{d\epsilon}\right | _{\eta = \eta_0} = \frac{\lambda}{a_0(T - T^{*}) + 3b\eta_0^2}
\end{eqnarray}
where $\eta_0$ is the induced relative change of resistivity of the sample at zero applied voltage comparing to before it was glued on the piezo. It is a function of temperature and was simultaneously measured when we performed voltage sweep at fix temperature in order to acquire $\eta(\epsilon)$. Representative data corresponding to the data of $d\eta/d\epsilon$ presented in Fig. \ref{Fig:Fig3} in the main text are shown in Fig. \ref{Fig:S3}. A numerical fit was performed based on the following expression:
\begin{eqnarray}
d\eta/d\epsilon = \frac{P_1}{T - P_2 + P_3\eta_0^2} + P_4
\end{eqnarray}
where $d\eta/d\epsilon$ and $\eta_0$ are functions of temperature $T$, and are simultaneously fitted by four parameters. $P_1$, $P_2$ and $P_3$ correspond to $\lambda/a_0$, $T^{*}$ and $3b/a_0$ in the Ginzburg-Landau theory respectively. $P_4 = \chi_0$ is a new parameter we introduced to take into account the intrinsic piezoresistive response of metals and semiconductors that occurs due to the induced orthorhombicity even in the absence of nematic fluctuations. The obtained fit parameters are shown in Fig. \ref{Fig:Fig4} and Fig. \ref{Fig:S4} as a function of cobalt concentration $x$. The contribution due to the built in strain is usually of the order of a few Kelvin, which will only be relevant if $T$ is very close to $T^*$.
As described in the main text, $T^{*}$ changes sign near optimal doping, which is consistent with an electronic nematic quantum phase transition. Inspection of Fig.\ref{Fig:S4}, also reveals an enhancement of $\lambda/a_0$ and $\chi_0$ near optimal doping, which can be consistently explained by an enhancement of electron-lattice coupling $\lambda$. The enhancement of $\lambda$ not only increases $\lambda/a_0$, but also could potentially affect the ``background" piezoresistivity effect measured by $\chi_0$. However the numerical fit also becomes less consistent close to optimal doping (Fig. \ref{Fig:Fig3}), which might be due to the effect of quantum critical fluctuations. To distinguish the effect of electron-lattice coupling and quantum critical fluctuations is beyond the scope of this paper.
\begin{figure}
\includegraphics[width=8.5cm]{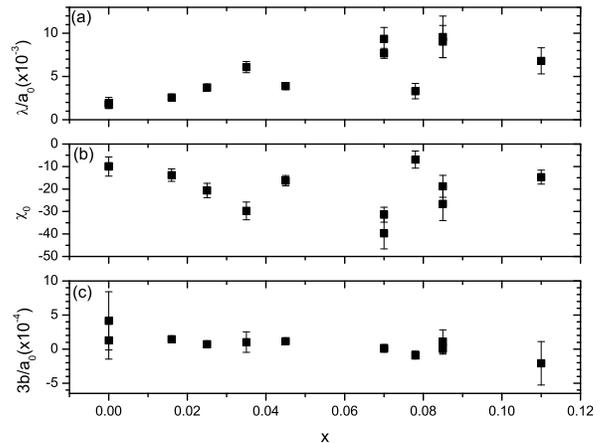}
\caption{\label{Fig:S4}Doping dependence of the parameters obtained from the numerical fit based on the Ginzburg-Landau theory.}
\end{figure}

Finally, the parameter $3b/a_0$ remains mostly constant across underdoped and optimal doped compositions but becomes negative for some overdoped compositions. This might be due to the fact that both $d\eta/d\epsilon$ and $\eta_0$ become small for overdoped compounds, which makes the fitting less constrained. On the other hand for the overdoped compounds there is no phase transition at finite temperature, and hence the sign of parameter $b$ is less crucial, and higher order coefficients can be introduced to bound the free energy.

\subsection{References}

S1. J.-H. Chu {\it et al}.,  Phys.\ Rev.\ B {\bf 79}, 014506 (2009))\\
S2. Part No. PSt 150/5×5×7, from Piezomechanik, Munich, Germany. \\
S3. Part No. 14250, General Purpose Adhesive Epoxy, from Devcon, U.S.A.\\
S4. Part No. WK-XX-062TT-350, General Purpose Strain Gages - Tee Rosette, from Vishay Precision Group.\\
S5. R. A. Cowley, Phys.\ Rev.\ B \textbf{13}, 4877 (1976)

\end{document}